\documentclass[11pt]{article}
\usepackage{times}
\usepackage{geometry}
\usepackage[utf8]{inputenc}
\usepackage{enumitem,amssymb}
\usepackage{ragged2e}
\usepackage{graphicx}
\usepackage{comment}
\usepackage{multicol}
\usepackage[usenames]{xcolor} 
\usepackage{wrapfig}

\definecolor{xlinkcolor}{cmyk}{1,1,0,0}
\usepackage{url}
\usepackage[
 colorlinks=true,    
 linkcolor=xlinkcolor,     
 citecolor=xlinkcolor,     
 filecolor=xlinkcolor,  
 urlcolor=xlinkcolor,      
 final=true
]{hyperref}
\usepackage[super,sort&compress]{natbib}
\usepackage{enumitem}
\setenumerate{itemsep=0mm}

\begin{document}
\begin{raggedright} 
\huge
Snowmass2021 - Letter of Interest \hfill \\[+1em]
\textit{Cosmology Intertwined I: Perspectives for the Next Decade} \hfill \\[+1em]
\end{raggedright}

\normalsize

\noindent {\large \bf Thematic Areas:}  (check all that apply $\square$/$\blacksquare$)

\noindent $\blacksquare$ (CF1) Dark Matter: Particle Like \\
\noindent $\square$ (CF2) Dark Matter: Wavelike  \\ 
\noindent $\square$ (CF3) Dark Matter: Cosmic Probes  \\
\noindent $\blacksquare$ (CF4) Dark Energy and Cosmic Acceleration: The Modern Universe \\
\noindent $\square$ (CF5) Dark Energy and Cosmic Acceleration: Cosmic Dawn and Before \\
\noindent $\square$ (CF6) Dark Energy and Cosmic Acceleration: Complementarity of Probes and New Facilities \\
\noindent $\blacksquare$ (CF7) Cosmic Probes of Fundamental Physics \\
\noindent $\blacksquare$ (Other) TF01, TF09 \\

\noindent {\large \bf Contact Information:}\\
Eleonora Di Valentino (JBCA, University of Manchester, UK) [eleonora.divalentino@manchester.ac.uk]\\

\noindent {\large \bf Authors:}  \\[+1em]
Eleonora Di Valentino (JBCA, University of Manchester, UK)\\
Luis A. Anchordoqui (City University of New York, USA)\\
\"{O}zg\"{u}r Akarsu (Istanbul Technical University, Istanbul, Turkey) \\
Yacine Ali-Haimoud (New York University, USA)\\
Luca Amendola (University of Heidelberg, Germany)\\
Nikki Arendse (DARK, Niels Bohr Institute, Denmark) \\
Marika Asgari (University of Edinburgh, UK)\\
Mario Ballardini (Alma Mater Studiorum Universit\`a di Bologna, Italy)\\
Spyros Basilakos (Academy of Athens and Nat. Observatory of Athens, Greece) \\
Elia Battistelli (Sapienza Universit\`a di Roma and INFN sezione di Roma, Italy)\\
Micol Benetti (Universit\`a degli Studi di Napoli Federico II and INFN sezione di Napoli, Italy)\\
Simon Birrer (Stanford University, USA)\\
Fran\c{c}ois R. Bouchet (Institut d'Astrophysique de Paris, CNRS \& Sorbonne University, France) \\
Marco Bruni (Institute of Cosmology and Gravitation, Portsmouth, UK, and INFN Sezione di Trieste, Italy)\\
Erminia Calabrese (Cardiff University, UK)\\
David Camarena (Federal University of Espirito Santo, Brazil) \\
Salvatore Capozziello (Universit\`a degli Studi di Napoli Federico II, Napoli, Italy) \\
Angela Chen (University of Michigan, Ann Arbor, USA)\\
Jens Chluba (JBCA, University of Manchester, UK)\\
Anton Chudaykin (Institute for Nuclear Research, Russia) \\
Eoin \'O Colg\'ain (Asia Pacific Center for Theoretical Physics, Korea) \\
Francis-Yan Cyr-Racine (University of New Mexico, USA) \\
Paolo de Bernardis (Sapienza Universit\`a di Roma and INFN sezione di Roma, Italy) \\
Javier de Cruz P\'erez (Departament FQA and ICCUB, Universitat de Barcelona, Spain)\\
Jacques Delabrouille (CNRS/IN2P3, Laboratoire APC, France \& CEA/IRFU, France \& USTC, China)\\
Jo Dunkley (Princeton University, USA)\\
Celia Escamilla-Rivera (ICN, Universidad Nacional Aut\'onoma de M\'exico, Mexico) \\
Agn\`es Fert\'e (JPL, Caltech, Pasadena, USA)\\
Fabio Finelli (INAF OAS Bologna and INFN Sezione di Bologna, Italy) \\
Wendy Freedman (University of Chicago, Chicago IL, USA)\\
Noemi Frusciante (Instituto de Astrof\'isica e Ci\^encias do Espa\c{c}o, Lisboa, Portugal)\\
Elena Giusarma (Michigan Technological University, USA) \\
Adri\`a G\'omez-Valent (University of Heidelberg, Germany)\\
Will Handley (University of Cambridge, UK) \\
Ian Harrison (JBCA, University of Manchester, UK) \\
Luke Hart (JBCA, University of Manchester, UK)\\
Alan Heavens (ICIC, Imperial College London, UK)\\
Hendrik Hildebrandt (Ruhr-University Bochum, Germany)\\
Daniel Holz (University of Chicago, Chicago IL, USA)\\
Dragan Huterer (University of Michigan, Ann Arbor, USA)\\
Mikhail M. Ivanov (New York University, USA) \\
Shahab Joudaki (University of Oxford, UK and University of Waterloo, Canada) \\
Marc Kamionkowski (Johns Hopkins University, Baltimore, MD, USA) \\
Tanvi Karwal (University of Pennsylvania, Philadelphia, USA) \\
Lloyd Knox (UC Davis, Davis CA, USA)\\
Suresh Kumar (BITS Pilani, Pilani Campus, India) \\
Luca Lamagna (Sapienza Universit\`a di Roma and INFN sezione di Roma, Italy) \\
Julien Lesgourgues (RWTH Aachen University) \\
Matteo Lucca (Universit\'e Libre de Bruxelles, Belgium)\\
Valerio Marra (Federal University of Espirito Santo, Brazil) \\
Silvia Masi (Sapienza Universit\`a di Roma and INFN sezione di Roma, Italy) \\
Sabino Matarrese (University of Padova and INFN Sezione di Padova, Italy) \\
Arindam Mazumdar (Centre for Theoretical Studies, IIT Kharagpur, India) \\
Alessandro Melchiorri (Sapienza Universit\`a di Roma and INFN sezione di Roma, Italy)\\
Olga Mena (IFIC, CSIC-UV, Spain)\\
Laura Mersini-Houghton (University of North Carolina at Chapel Hill, USA) \\
Vivian Miranda (University of Arizona, USA) \\
Cristian Moreno-Pulido (Departament FQA and ICCUB, Universitat de Barcelona, Spain)\\
David F. Mota (University of Oslo, Norway) \\
Jessica Muir (KIPAC, Stanford University, USA)\\
Ankan Mukherjee (Jamia Millia Islamia Central University, India) \\
Florian Niedermann (CP3-Origins, University of Southern Denmark) \\
Alessio Notari (ICCUB, Universitat de Barcelona, Spain) \\
Rafael C. Nunes (National Institute for Space Research, Brazil)\\
Francesco Pace (JBCA, University of Manchester, UK)\\
Andronikos Paliathanasis (DUT, South Africa and UACh, Chile) \\
Antonella Palmese (Fermi National Accelerator Laboratory, USA) \\
Supriya Pan (Presidency University, Kolkata, India)\\
Daniela Paoletti (INAF OAS Bologna and INFN Sezione di Bologna, Italy)\\
Valeria Pettorino (AIM, CEA, CNRS, Universit\'e Paris-Saclay, Universit\'e de Paris, France) \\
Francesco Piacentini (Sapienza Universit\`a di Roma and INFN sezione di Roma, Italy)\\
Vivian Poulin (LUPM, CNRS \& University of Montpellier, France) \\
Marco Raveri (University of Pennsylvania, Philadelphia, USA) \\
Adam G. Riess (Johns Hopkins University, Baltimore, USA) \\
Vincenzo Salzano (University of Szczecin, Poland)\\
Emmanuel N. Saridakis (National Observatory of Athens, Greece)\\
Anjan A. Sen (Jamia Millia Islamia Central University New Delhi, India) \\
Arman Shafieloo (Korea Astronomy and Space Science Institute (KASI), Korea)\\
Anowar J. Shajib (University of California, Los Angeles, USA) \\
Joseph Silk (IAP Sorbonne University \& CNRS, France, and Johns Hopkins University, USA)\\
Alessandra Silvestri (Leiden University, NL)\\
Martin S. Sloth (CP3-Origins, University of Southern Denmark) \\
Tristan L. Smith (Swarthmore College, Swarthmore, USA)\\ 
Joan Sol\`a Peracaula (Departament FQA and ICCUB, Universitat de Barcelona, Spain)\\
Carsten van de Bruck (University of Sheffield, UK) \\
Licia Verde (ICREA, Universidad de Barcelona, Spain)\\
Luca Visinelli (GRAPPA, University of Amsterdam, NL) \\
Benjamin D. Wandelt (IAP Sorbonne University \& CNRS, France, and CCA, USA) \\
Deng Wang (National Astronomical Observatories, CAS, China) \\
Jian-Min Wang (Key Laboratory for Particle Astrophysics, IHEP of the CAS, Beijing, China) \\
Anil K. Yadav (United College of Engg. \& Research, GN, India)\\
Weiqiang Yang (Liaoning Normal University, Dalian, China) \\

\noindent {\large \bf Abstract:} 
The standard $\Lambda$ Cold Dark Matter cosmological model provides an amazing description of a wide range of astrophysical and astronomical data. However, there are a few big open questions, that make the standard model look like a first-order approximation to a more realistic scenario that still needs to be fully understood. In this Letter of Interest we will list a few important goals that need to be addressed in the next decade, also taking into account the current discordances present between the different cosmological probes, as the Hubble constant $H_0$ value, the $\sigma_8 – S_8$ tension, and the anomalies present in the Planck results. Finally, we will give an overview of upgraded experiments and next-generation space-missions and facilities on Earth, that will be of crucial importance to address all these questions.

\clearpage

\noindent {\bf The big questions and goals for the next decade --} The
standard $\Lambda$ Cold Dark Matter ($\Lambda$CDM) cosmological model
provides an amazing description of a wide range of astrophysical and
astronomical data. Over the last few years, the parameters governing
$\Lambda$CDM have been constrained with unprecedented accuracy by
precise measurements of the cosmic microwave background
(CMB)~\cite{Akrami:2018vks,Aghanim:2018eyx}. However, despite its incredible
success, $\Lambda$CDM still cannot explain key concepts in our
understanding of the universe, at the moment based on unknown quantities like Dark Energy (DE), Dark Matter (DM) and Inflation. Therefore, in the next decade the first challenges would be to answer the following questions:
\begin{itemize}[noitemsep,topsep=0pt]
\item What is the nature of dark energy and dark matter?
\item Did the universe have an inflationary period? How did it happen? What is the level of non-gaussianities?
\item Does gravity behave like General Relativity even at horizon size scales? Is there Modified Gravity?
\item Do we need quantum gravity, or an unified theory for quantum field theory and General Relativity?
\item Is the universe flat or closed?
\item What is the age of the universe?
\item Do we actually need physics beyond the Standard Model (SM) of particle physics? 
\item For each elementary particle, there is an antiparticle that has exactly the very same properties but opposite charge. Then, why
 we do not see antimatter in the universe?
\item Will the swampland conjectures within string theory help with fine-tuning problems in cosmology? Alternatively, will cosmology help us observationally test conjectures from string theory?
  \end{itemize}
  
\noindent The $\Lambda$CDM model can therefore be seen as an approximation to a more realistic scenario that still needs to be fully understood. However, since the $\Lambda$CDM model provides an extremely good fit of the data, deviations from the model are not expected to be too drastic from the phenomenological point of view, even if they can be conceptually really different. In particular, discrepancies with different statistical significance developing between observations at early and late cosmological time may involve the addition of new physics ingredients~\cite{Verde:2019ivm} in the $\Lambda$CDM minimal model. For this reason, it is timely to investigate the disagreement at more than $4\sigma$ about the Hubble constant $H_0$~\cite{DiValentino:2020zio}, followed by the tension at $\sim 3\sigma$ on $\sigma_8 - S_8$~\cite{DiValentino:2020vvd}, and the anomalies in the Planck experiment results about the excess of lensing, the curvature of the Universe or its age~\cite{DiValentino:2020srs}. In the next decade we aim to address these discrepancies solving the following key questions:
  
\begin{itemize}[noitemsep,topsep=0pt]
\item What is the origin of the sharpened tension in the observed and
  inferred values of $H_0$, $f\sigma_8$, and $S_8$?
 \item Is it possible that some portion (with an outside chance of all)
  of the tension may still be systematic
  errors in the current measurements?
\item Is the tension a statistical fluke or is it pointing to new physics?
  \item Is it possible to explain the  tension without changing the
  standard $\Lambda$CDM cosmology?
  \item Is there an underlying new physics that can accommodate this tension? 
  \end{itemize}

\noindent In order to address all the open questions, and to change the $\Lambda$CDM from an effective model to a physical model, testing the different predictions, the goals for the next decade will be to:
\begin{itemize}[noitemsep,topsep=0pt]
\item improve our understanding of systematic uncertainties;
\item maximize the amount of information that can be extracted from the data by considering new analysis frameworks and exploring alternative connections between the different phenomena;
\item improve our understanding of the physics on non-linear scales;
\item de-standardize some of the $\Lambda$CDM assumptions, or carefully label them in the survey analysis pipelines, to pave the road to the beyond-$\Lambda$CDM models tests carried out by different groups. 
\end{itemize}
This agenda is largely achievable in the next decade, thanks to a coordinated effort from the side of theory, data analysis, and observation. In separate LoI's~\cite{DiValentino:2020zio,DiValentino:2020vvd,DiValentino:2020srs} we provide a thorough discussion of these challenging questions, showing also the impossibility we have at the moment of solving all the tensions at the same time.

\noindent {\bf Stepping up to the new challenges --} The next decade
will provide a compelling and complementary view of the cosmos through a combination of enhanced statistics,
refined analyses afforded by upgraded experiments and next-generation
space-missions and facilities on Earth:
\begin{itemize}[noitemsep,topsep=0pt]
\item Local distance ladder observations will achieve a precision in the $H_0$ measurement of 1\%~\cite{Riess:2020xrj}.
\item Gravitational time delays will reach a $\sim 1.5\%$ precision on $H_0$ without relying on assumption on the radial mass density profiles~\cite{Birrer:2020jyr} with resolved stellar kinematics measurement from JWST or the next generation large ground based extremely large telescopes (ELTs).
\item CMB-S4 will constrain departures from the thermal history of the universe predicted by the SM~\cite{Abazajian:2019eic,Abazajian:2019tiv}. The departures are usually conveniently quantified by the contribution of light relics to the effective number of relativistic species in the early Universe, $N_{\rm eff}$~\cite{Steigman:1977kc}. CMB-S4 will constrain $\Delta N_{\rm eff} \leq 0.06$ at the 95\% confidence level allowing detection of, or constraints on, a wide range of light relic particles even if they are too weakly interacting to be detected by lab-based experiments~\cite{Abazajian:2019eic}.
\item The Euclid space-based survey mission~\cite{Laureijs:2011gra} will use cosmological probes (gravitational lensing, baryon acoustic oscillations (BAO) and galaxy clustering) to investigate the nature of DE, DM, and gravity~\cite{Capozziello:2011et}.
\item The Rubin Observatory Legacy Survey of Space and Time (LSST~\cite{Ivezic:2008fe}) is planned to undertake a 10-year survey beginning in 2022. LSST will chart 20 billion galaxies, providing multiple simultaneous probes of DE, DM, and $\Lambda$CDM~\cite{Abell:2009aa,Zhan:2017uwu,Sahni:2006pa}.
\item The Roman Space Telescope (formerly known as WFIRST~\cite{akeson2019wide}) will be hundreds of times more efficient than the Hubble Space Telescope, investigating DE, cosmic acceleration, exoplanets, cosmic voids.
\item The combination of LSST, Euclid, and WFIRST will improve another factor of ten the cosmological parameter bounds, allowing us to distinguish between models candidates to alleviate the tensions. 
\item The Square Kilometre Array (SKA) will be a multi-purpose radio-interferometer, with up to 10 times more sensitivity, and 100 times faster survey capabilities than current radio-interferometers, providing leading edge science involving multiple science disciplines. SKA will be able to probe DM properties (interactions, velocities and nature) through the detection of the redshifted 21 cm line in neutral hydrogen (HI), during the so-called Dark Ages, before the period of reionization. SKA will also be able to test the DE properties and the difference between some MG and DE scenarios by detecting the $21$~cm HI emission line from around a billion galaxies over 3/4 of the sky, out to a redshift of $z \sim 2$.
\item CMB spectral distortions will be a possible avenue to test a variety of different cosmological models in the next decade \cite{Chluba:2019kpb}, with applications ranging from non-standard inflationary scenarios and beyond the SM physics \cite{Chluba:2019nxa} to the $H_0$ tension \cite{Abitbol:2019ewx, Lucca:2020fgp} (see also~\cite{Kogut:2019vqh,Lucca:2019rxf} for recent reviews);
\item $O(10^5)$ voids will be detected in upcoming surveys, that can place constraints on the expansion history of the universe~\cite{Hamaus:2020cbu} following a purely geometric approach, and distinguish different gravity models~\cite{Zhang:2020qkd}.
\item Gravitational wave (GW) coalescence events would provide a precise measurement of $H_0$~\cite{Schutz:1986gp,Abbott:2017xzu}. The LIGO-Virgo network operating at design sensitivity is expected to constrain $H_0$ to a precision of $\sim 2\%$ within 5 years and $1\%$ within a decade~\cite{Chen:2017rfc}. Moreover, in~\cite{Mukherjee:2018ebj} it is shown that even in absence of electromagnetic counterpart, it is possible to measure $H_0$ cross-correlating with a clustering tracer, as a galaxy survey. Therefore, black hole binaries should provide a competitive $H_0$ estimate faster~\cite{Mukherjee:2020hyn}.
\item CERN's LHC experiments ATLAS and CMS will provide complementary information by searching for the elusive DM particle and hyperweak gauge interactions of light relics~\cite{Buchmueller:2017qhf,Penning:2017tmb,CidVidal:2018eel,Anchordoqui:2020znj}. In addition, the ForwArd Search ExpeRiment (FASER) will search for light hyperweakly-interacting particles produced in the LHC’s high-energy collisions in the far-forward region~\cite{Ariga:2018pin,Ariga:2019ufm,Feng:2019bci}.
\end{itemize}   

\noindent Concluding, the current present tensions and discrepancies among different measurements, in particular the $H_0$ tension as the most significant one, offer crucial insights in our understanding of the universe. For example, the standard distance ladder result has many steps in common with the accelerating universe discovery (which gave cosmology the evidence for DE). So, whatever the definite finding may be, whether about stars and their evolution, or DE, this is going to have far reaching consequences.
\clearpage

\bibliographystyle{utphys}
\bibliography{perspectives}
\vspace{3.5in}


\end{document}